\documentclass[%
superscriptaddress,
nofootinbib,
 amsmath,amssymb,
 aps,
 notitlepage,
 prapplied,
 twocolumn,
 longbibliography,
]{revtex4-1}

\usepackage{graphicx}
\usepackage{dcolumn}
\usepackage{bm}
\usepackage[colorlinks,
    citecolor=Blue,
    linkcolor=Maroon,
    urlcolor=ForestGreen,
    ]{hyperref}
\usepackage{siunitx}

\usepackage[capitalise,nameinlink]{cleveref}

\usepackage[usenames,dvipsnames,table]{xcolor}
\usepackage{easy-todo}

\newcommand{\cNA}[1]{{\color[RGB]{250,0,0}#1}}

\newcommand{\Number}[1]{{\color[RGB]{0,0,0}#1}}
\newcommand{\opendata}[1]{}

\newcommand{\NUCFPCIERA}{{Center for Fundamental Physics (CFP), Center for Interdisciplinary and Exploratory Research in Astrophysics (CIERA), Department of Physics, Northwestern University, Evanston, USA}}

\newcommand{\PTB}{Physikalisch-Technische Bundesanstalt, Berlin, Germany}

\newcommand{\NU}{{ Center for Fundamental Physics,Department of Physics and Astronomy, Northwestern University}}

\newcommand{\IU}{Department of Physics, Indiana University, 727 E Third St, Bloomington, IN 47405}

\newcommand{\SUB}{Department of Physics and Applied Physics, Stanford University, 382 Via Pueblo, Stanford, CA 94305}

\newcommand{\CAPP}{Center for Axion and Precision Physics Research, IBS, 193 Munji-Ro, Daejeon, South Korea, 34051}
\newcommand{\KAIST}{Department of Physics, KAIST, 291 Daehak-Ro, Daejeon, South Korea, 34141}

\newcommand{\KRISS}{KRISS, 267 Gajeong-ro, Yuseong-gu, Daejeon 34113, Republic of Korea}

\begin{abstract}
	ARIADNE is a nuclear-magnetic-resonance-based experiment that will search for novel axion-induced spin-dependent interactions between an unpolarized source mass rotor and a nearby sample of spin-polarized $^3$He gas. To detect feeble axion signals at the sub-atto-Tesla level, the experiment relies on low magnetic background and  noise. We measure and characterize the magnetic field background from a prototype tungsten rotor.  We show that the requirement is met with our current level of tungsten purity and demagnetization process. We further show that the noise is dominantly caused by a few discrete dipoles, likely due to a few impurities trapped inside the rotor during manufacturing. This is done via a numerical optimization pipeline which fits for the locations and magnetic moments of each dipole. We find that under the current demagnetization, the magnetic moment of trapped impurities is bounded at \Number{\SI{1e-9}{\ampere\meter\squared}}.
\end{abstract}

\begin{document}

\preprint{APS/123-QED}

\title{Characterization of magnetic field noise in the ARIADNE source mass rotor}

\author{Nancy Aggarwal}
\email{nancy.aggarwal@northwestern.edu}
\affiliation{\NUCFPCIERA}


\author{A. Schnabel}
\affiliation{\PTB}
\author{J. Voigt}
\email{Jens.Voigt@ptb.de}
\affiliation{\PTB}
\author{Alex Brown}
\email{deceased}
\affiliation{\IU}
\author{J.C. Long}
\affiliation{\IU}
\author{L. Trahms}
\affiliation{\PTB}
\author{A. Fang}
\affiliation{\SUB}
\author{A.A. Geraci}
\affiliation{\NU}
\author{A. Kapitulnik}
\affiliation{\SUB}
\author{D. Kim}
\affiliation{\CAPP}
\affiliation{\KAIST}
\author{Y. Kim}
\affiliation{\CAPP}
\affiliation{\KAIST}
\author{I. Lee}
\affiliation{\IU}
\author{Y.H. Lee}
\affiliation{\KRISS}
\author{C.Y. Liu}
\affiliation{\IU}
\author{C. Lohmeyer}
\affiliation{\NU}
\author{A. Reid}
\affiliation{\IU}
\author{Y. Semertzidis}
\affiliation{\CAPP}
\affiliation{\KAIST}
\author{Y. Shin}
\affiliation{\CAPP}
\author{J. Shortino}
\affiliation{\IU}
\author{E. Smith}
\affiliation{\IU}
\author{W.M. Snow}
\affiliation{\IU}
\author{E. Weisman (ARIADNE Collaboration)}
\affiliation{\NU}

\date{\today}

\maketitle


\section{Introduction}

	The QCD axion is a particle predicted to exist in order to explain the strong-CP problem, 
connected with the lack of CP-violation observed in the strong interactions \cite{PQaxion,Weinber_boson,Wilczek_instaton,Moody_spin_dependent_forces}. Experimental searches for a neutron electric dipole moment constrain the angle $\theta_{\rm{QCD}}$ which  would be expected to be of order \SI{1}{\radian}, to be less than \SI{1e-10}{\radian} \cite{mercury,neutronEDM2020}. The axion provides a dynamical mechanism to explain the unnatural smallness of this angle. In addition, the axion is an excellent candidate to explain the dark matter in our universe. 

The axion can 
mediate novel short-range spin dependent forces between fermions. 
	The Axion Resonant InterAction DetectioN Experiment (ARIADNE) will search for a spin-dependent interaction that is mediated by the QCD axion between a gas of hyper-polarized $^3$He nuclear spins and an unpolarized source mass \cite{Ariadneproposal,Ariadneprogress}.
	Since the axion field interacts with the $^3$He gas via a spin-dependent interaction,  the axion field can be treated as a fictitious transverse magnetic field. This effect can be resonantly enhanced by modulating the unpolarized mass at the Larmor frequency of the $^3$He gas. In order to distinguish the axion signal
	, real magnetic fields need to be shielded to a level below the expected fictitious magnetic field.
	In  ARIADNE, a superconducting niobium magnetic shield will enclose the $^3$He sample, separating the $^3$He sample from the source mass, and shielding the \(^3\)He gas from standard-model field perturbations. 
	
	In the experiment, the modulation of the unpolarized mass will be implemented by rotating a Tungsten wheel that has eleven 
	cutouts around its circumference. The modulation frequency ($f_\mathrm{mod}$ = \SI{55}{\hertz} to \SI{99}{\hertz}) will then be eleven times the rotation frequency ($f_\mathrm{rot}$ = \SI{5}{\hertz} to \SI{9}{\hertz}).
	The sub-attoTesla axion-induced effective magnetic field created by this mass is at $f_\mathrm{mod}$ and is not shielded by the superconductor. 
	The magnetic background from the rotor and the shielding factor of the superconducting shield together play a combined role in achieving the target axion sensitivity. 
	For instance, assuming a baseline shielding factor of \Number{\SI{1e8}{}}, any coherent magnetic field background at the modulation frequency must be lower than \Number{\SI{1e-12}{\tesla}}  in order for ARIADNE to operate at a sensitivity of \Number{\SI{1e-20}{\tesla}}. 
	Additionally, assuming an integration time of \Number{\SI{1e6}{\second}} and the same shielding factor, the magnetic field noise from stochastic processes (e.g. Johnson noise) must be lower than \Number{\SI{1}{\nano\tesla\per\sqrt{Hz}}}. From this point, we will refer something as \textit{noise} if its effect can be lowered by integrating for longer times, and as \textit{background} if it does not get suppressed by integrating longer.

Several relevant sources of magnetic field noise and potential systematic backgrounds 
	 in ARIADNE are associated with the magnetism of the source mass rotor itself.  In this paper, we describe optical magnetometry measurements performed on the prototype tungsten source mass rotor to be used in the experiment. We study the frequency dependence of magnetic noise generated as the rotor rotates within a magnetically shielded environment, and we show that the design requirements are met with our current level of tungsten purity and demagnetization process.  We further show that the magnetic field profile from the rotor suggests the presence of a few discrete dipoles, likely due to impurities trapped inside the rotor during manufacturing.  

Tungsten is often a material of a choice for tests of exotic spin- and/or mass-dependent interactions which couple to nucleons, given its high elastic modulus and high-nucleon density. Tungsten has also been used in tests for novel gravitational interactions at sub-millimeter range \cite{weld2008}, where magnetic field due to magnetic impurities or impurities of superconducting material deposits can be a significant concern. Gold is occasionally used as well for its high density. However, gold is a soft metal, making it more challenging to machine.  Thus this work is also relevant for other existing and planned ``fifth-force" experiments for novel exotic interactions.

\section{Expected magnetic noise sources}


The fundamental noise source which will limit ARIADNE's sensitivity is the quantum noise caused by the transverse spin-projection in the ${^3}$He. For the target polarization level and nucleus density, this noise is expected to be around \SI{3e-19}{\tesla\per\sqrt{Hz}} \((1000\, \mathrm{s}/T_2)^{1/2}\), where \(T_2\) is the \(^3\)He spin-relaxation time \cite{Ariadneproposal}. 

Besides the $^3$He sample, the source mass produces numerous magnetic noises and backgrounds which must be lowered below the spin-projection noise so as to be spin-projection limited. 
We summarize the various noise and background requirements from the tungsten rotor \cite{Ariadneproposal}, assuming a baseline shielding factor of \SI{1e8}{}, integration time of \SI{1e6}{\second}, and an axion sensitivity of \SI{1e-20}{\tesla}. A higher shielding factor or longer integration times can be targeted to reduce the effect of source mass and/or to probe deeper into the axion parameter space.



At the operating temperature of \Number{\SI{170}{\kelvin}} and at the modulation frequency \(f_\mathrm{mod}\) the contribution from thermal (Johnson) noise is expected to be around \Number{\SI{1e-12}{\tesla\per\sqrt{Hz}}} \cite{Ariadneproposal}. After shielding this will become \Number{\SI{1e-20}{\tesla\per\sqrt{Hz}}}, or \Number{\SI{1e-23}{\tesla}} with integration.

Since the Tungsten source mass is uncharged and being spun around its axis, it will acquire a magnetization due to the Barnett effect \cite{barnett}. For rotation frequencies from \Number{\SI{5}{Hz}} to \Number{\SI{9}{Hz}}, we expect the field from this magnetization to be around \Number{\SI{1e-14}{\tesla}} \cite{Ariadneproposal}, suppressed to \Number{\SI{1e-22}{T}} with the shield.


Another source of magnetic backgrounds can be caused by magnetic impurities inside the tungsten source mass. To estimate this effect, we assume iron atoms trapped inside the source mass at a 1 ppm purity level. The magnetic field from such impurities is calculated in Ref. \cite{Ariadneproposal} to be between \SI{1e-17}{\tesla} and \SI{1e-9}{\tesla}, which can be shielded to  \SI{e-25}{T} -- \SI{e-17}{T}. The range comes from the distribution of the alignment of the Fe spins. This field at 1 ppm impurity level is outside the specification if the spins are near-perfect aligned. One of the aims of this study is to determine the typical magnetic impurity distributions in the prototype tungsten rotor. Effects of residual magnetism due to hysteresis from magnetization-demagnetization cycles were studied in a previous publication and found to meet the requirements \cite{lohmeyer2020source}. 

 Finally, the intrinsic magnetic susceptibility of tungsten can lead to induced dipole moment for a given background field. The rotation of the source mass can lead to a modulation of magnetic field from the induced moment at the Larmor frequency. For a background field of \Number{\SI{1e-10}{\tesla}}, we expect a modulation with an amplitude of \Number{\SI{1e-14}{\tesla}} \cite{Ariadneproposal}, which can be shielded to \Number{\SI{1e-22}{T}} by the superconducting shield. This estimate can be used to assess how stringent the requirement on the background magnetic field in the experiment must be. For example, with these experimental parameters, for conducting an axion search at the \SI{1e-20}{\tesla} level, the background magnetic field thus should be maintained below the 10 nT level at the location of sprocket. This can be implemented in the experiment by using a combination of magnetic shielding and magnetic shim coils. Furthermore, any effect due to the magnetic susceptiblity of Tungsten can also be distinguished in the final ARIADNE experiment by varying the value of the background magnetic field: any axion-generated fictitious magnetic field will not depend on this background field whereas the effect due to the magnetic susceptibility has a linear dependence. 
 
%
%


\section{Experimental Setup}

\begin{figure}
		\includegraphics[width=\linewidth]{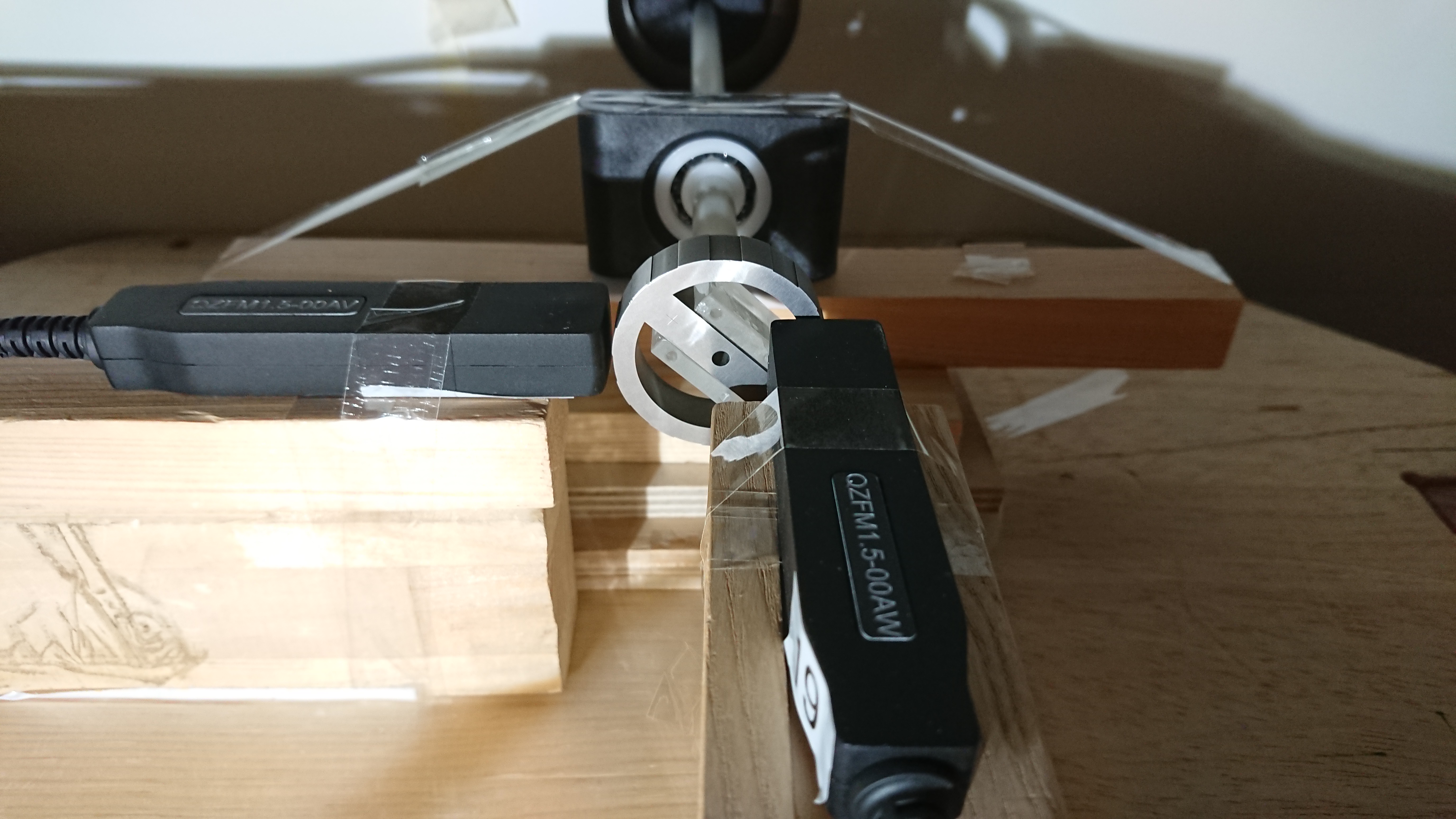}
	\hspace{0.04cm}
		\includegraphics[width=\linewidth]{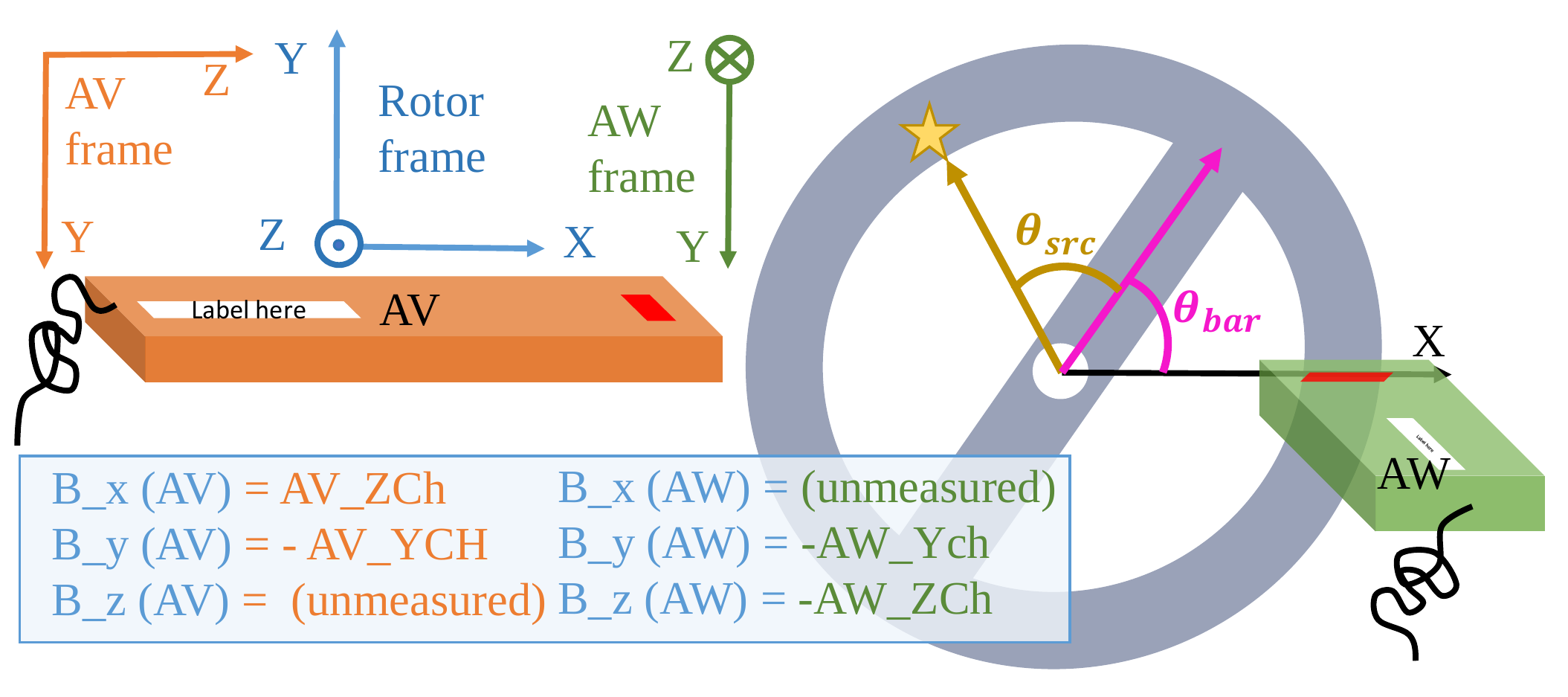}
	\caption{(upper) Photograph and (lower) schematic of the setup. Two QuSpin OPMs, AV (orange) and AW (green) are placed in proximity to the rotating ARIADNE source mass. The source mass has \Number{\SI{200}{\micro\meter}} cutouts along its circumference as can be seen in the photograph (not shown in the schematic). Each sensor measures its respective Y and Z direction. Their conversion to a unified lab-frame (blue) is also shown. \(\theta_\mathrm{bar}\) denotes the angular location of the central bar on the source mass at the beginning of the measurement. \(\theta_\mathrm{src}\) denotes the location of an example impurity in reference to the bar.}
	\label{fig:Setup}
\end{figure}

Our experimental setup consists of two optically pumped magnetometers (OPMs) placed in proximity to the source mass rotor which is mounted on a motor.

	The source mass has a height $1$ cm, inner diameter $3.4$ cm, and outer diameter $3.8$ cm, divided into $22$ sections of length $5.4$ mm. The section radii are modulated by approximately $200$ \textmu m in order to generate a time-varying potential at frequency $f_\mathrm{mod} = 11 \,f_{\rm{rot}}$. The factor of $11$ difference between $f_{\rm{rot}}$ and $f_\mathrm{mod}$ decouples mechanical vibration from the signal of interest. Tungsten that was more than $99.95\% $ pure \cite{midwesttungsten} was used to machine the source mass using wire electrical discharging machining \cite{wireEDM}. 
To keep the axion signal at a constant level, the tolerance of the cutouts is chosen to be $\pm 10$ \textmu m. The source mass height is chosen to maximize overlap with the ${^3}$He sample region. The smoothness of the top and bottom flat surfaces was achieved by finishing them with 220 and 440 grit abrasive sheets. The surface roughness was then measured using profilometry, and found to be $1.4$-$2.1$ \textmu m. 

The rotor is 
demagnetized using a hard-drive eraser (Verity Systems, bulk degausser SV 91m). It is then set up for rotation on an axle inside the Berlin magnetically shielded room-2 (BMSR-2) driven by an outside motor. In the ARIADNE experiment the sprocket will be driven by a Ti$_6$Al$_4$V shaft and precision ceramic bearings. For this study the rotor was affixed to a glass fiber reinforced plastic shaft, held by a commercial glass/plastic bearing..
 The two OPMs are placed next to it. These QZFM Gen 1.5 OPMs manufactured by QuSpin have demonstrated measurements at the \si{\pico\tesla} scale in a zero background field \cite{quspin}. 

\begin{figure}
	\includegraphics[width=0.5\textwidth]{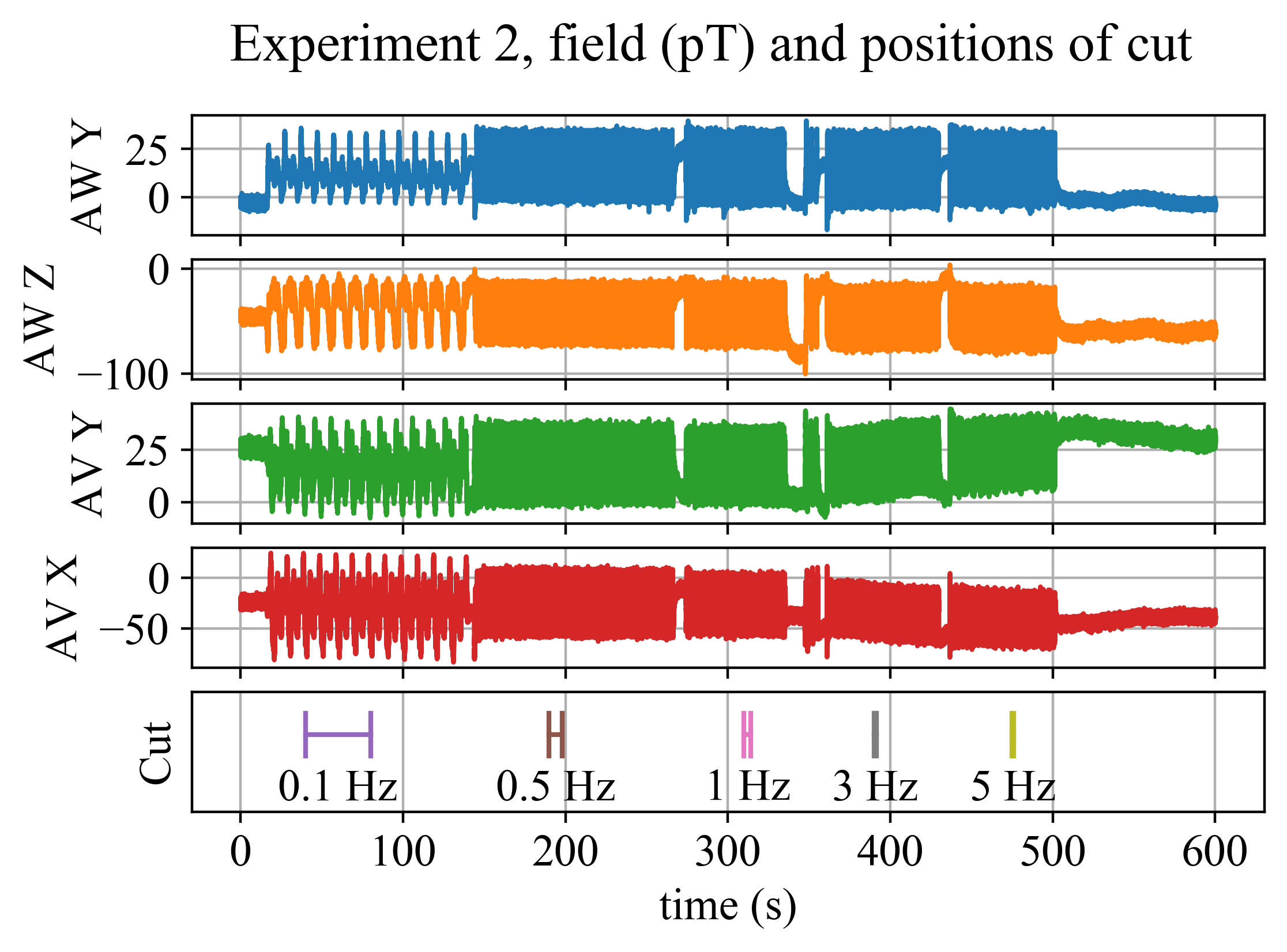}
	\caption{Raw data from the magnetometers AW and AV, in Y and Z direction for AW, and in X and Y directions for AV. The rotation is started at the slowest speed (\SI{0.1}{Hz}) at around 20 seconds, and the speed is then increased in steps. In the bottom panel, we show the locations of the cut to extract a few cycles at each frequency for the time-domain analysis.}
	\label{fig:data_cut_place}
\end{figure}

A photograph of the setup and a schematic is shown in \cref{fig:Setup}. The sensor to the left of the rotor has been named AV, and the sensor in front of the rotor has been named AW. The OPM in the sensor is located at the red mark shown in the schematic.

\begin{figure}
	\includegraphics[width=0.5\textwidth]{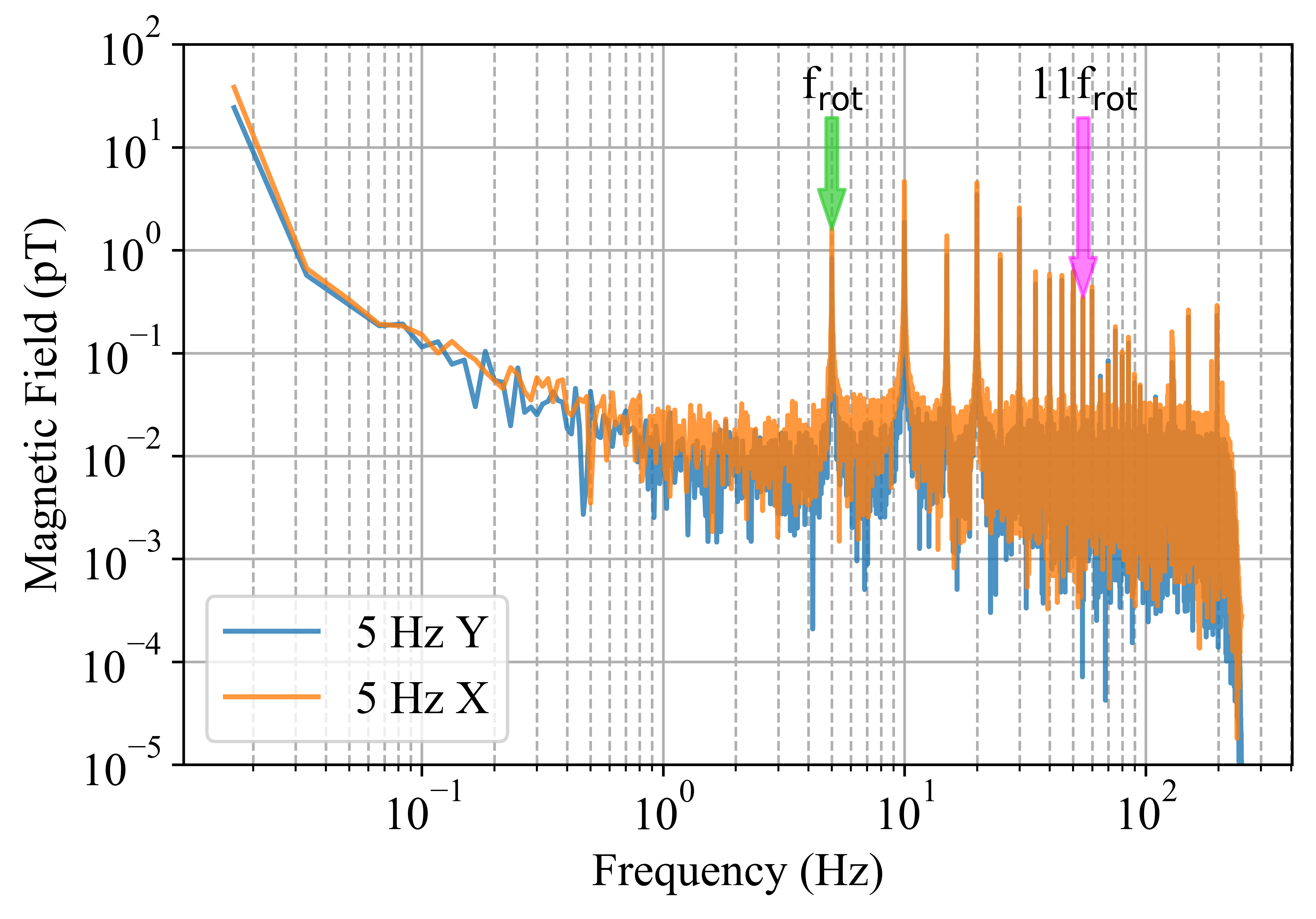}
	\caption{Frequency spectrum of the magnetic field measured by AV sensor in the X and Y directions, with the rotor being rotated at 5 Hz.}
	\label{fig:FFT_AV_5Hz}
\end{figure}

The OPM measures the field at its location in two axes, depicted by the orange and green axes in the figure. In the lab-frame, the sensor AW measures in the Y and Z direction, and the sensor AV measures in the X and Y directions.
The magnetic field at the sensors is sampled at 500 Hz and is recorded at five rotation frequencies: \SI{0.1}{Hz}, \SI{0.5}{Hz}, \SI{1}{Hz}, \SI{3}{Hz}, \SI{5}{Hz}. The full data is shown in \cref{fig:data_cut_place}. The data is analyzed both in time-domain as well as frequency domain, each providing separate information.


\section{Data Analysis}


 First we perform a spectral analysis to directly measure the magnetic field  amplitude at the modulation frequency. This is done for both axes of both sensors, and for all rotation frequencies. As an example, we show the fast Fourier transform (FFT) of the field measured by the sensor AV when the rotor is rotated at \SI{5}{\hertz} in \cref{fig:FFT_AV_5Hz}. We summarize the field amplitude for all other frequencies as well as the AW sensor in \cref{tb:FFTAmps}. We find that the amplitude of magnetic field at the modulation frequency is below \SI{1}{\pico\tesla} in all cases. In particular, the X axis of AV measures the magnetic field in the radial direction, which is the most relevant background source for the axion measurement.  The rotor in its current state thus meets the previously stated background requirements.
 
 From the same spectral analysis, we can also obtain a rough upper limit on the magnetic field noise from the source mass. We can conservatively assume that the entire amplitude is caused by magnetic field noise and convert it to a spectral density. For the rotation speed of \SI{5}{\hertz}, the measurement duration is \(\sim\) \SI{50}{\second}, giving a spectral density of roughly \SI{2}{\pico\tesla\per\sqrt{\hertz}} at the modulation frequency, \SI{55}{\hertz}. This conservative estimate meets the requirements for magnetic field noise.
 
 \begin{table}
 \begin{center}
 {\rowcolors{2}{black!20!}{black!10}
  	\begin{tabular}{|>{\centering\arraybackslash}p{1.6cm}||>{\centering\arraybackslash}p{1.2cm}|>{\centering\arraybackslash}p{1.2cm}|>{\centering\arraybackslash}p{1.2cm}|>{\centering\arraybackslash}p{1.2cm}|}
 	\hline
 	Frequency (Hz)& AVX (pT) & AVY (pT) & AWY (pT) & AWZ (pT) \\
 	\hline \hline
 	0.1 & 5.0 & 2.8 & 1.9 & 11.3\\
 	1.1 & 0.7 & 0.5 & 0.2 & 0.2\\
 	\hline
 	0.5 & 3.2 & 1.4 & 2.9 & 8.9 \\
 	5.5 & 0.5 & 0.4 & 0.3 & 0.2 \\
 	\hline
 	1 & 2.7 & 1.1 & 3.2 & 8.8 \\
 	11 & 0.3 & 0.3 & 0.3 & 0.2\\
 	\hline
 	3 & 1.3 & 0.6 & 2.1 & 5.1\\
 	33 & 0.2 & 0.2 & 0.2 & 0.2\\
 	\hline
 	5 & 1.6 & 0.8 & 2.9 & 6.8\\
 	55 &0.3 & 0.3 & 0.3 & 0.3\\
 	\hline
 	\end{tabular}
 	}
 \caption{Magnetic field amplitude at each sensor's location for both measurement axes. This amplitude has been obtained by converting the measurement to the frequency domain. For each row-pair, the amplitude at the rotation frequency \(f_\mathrm{rot}\) is shown in the top row (light gray), and at the modulation frequency \(f_\mathrm{mod} = 11 f_\mathrm{rot} \) is shown in the bottom row (dark gray). }
 \label{tb:FFTAmps}
 \end{center}
 \end{table}


Although the frequency-domain shows that the requirement for installing this rotor in the ARIADNE experiment is met, it is not suitable to distinguish various noise/background sources. Specifically, it is unclear if this field is explained by bulk noise mechanisms like Johnson noise, or a vibration noise, or by trapped magnetic impurities. 

 In the time-domain, we investigate whether the data can be significantly explained by a few isolated magnetic impurities inside the rotor. We do this by assuming the impurities act as magnetic dipoles and fitting the locations and moments of such dipoles to the data. 
 
\begin{figure}
 	\includegraphics[width=\linewidth]{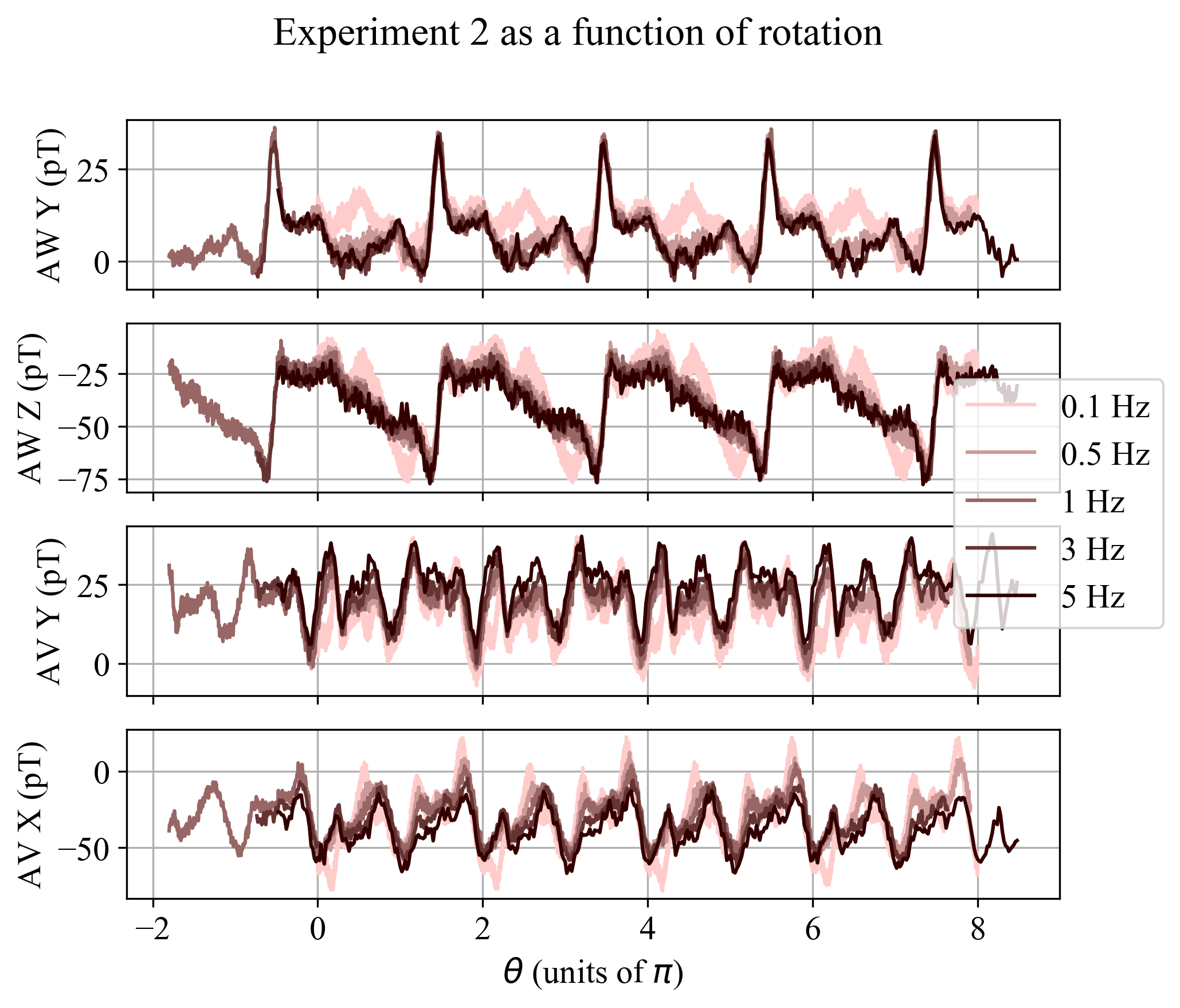}
 	\caption{Few cycles of data at all the rotation frequencies for both sensors and axes. Data for each frequency has been given a fixed shift in \(\theta\) to make them line up.}
 	\label{fig:cut_data}
 \end{figure}

 For the time-domain analysis, the data at each frequency is cut down and then averaged. Due to drifts in the rotation speed, the averaging was done over \(\sim 4\) cycles. The locations of cuts are shown in the bottom panel in \cref{fig:data_cut_place}. After performing the cuts, the data from multiple frequencies can be compared by converting the time axis to a rotation angle. This is shown in \cref{fig:cut_data}, where we have added arbitrary time offsets to line up the measurements from different rotation frequencies. We see that all measurements but the one at \SI{0.1}{Hz} are similar. We defer the analysis for the cause of deviation of \SI{0.1}{Hz} measurement to future work, especially since this speed is too low to be relevant for the axion measurement.
 
 \begin{figure*}
 	\includegraphics[width=\linewidth]{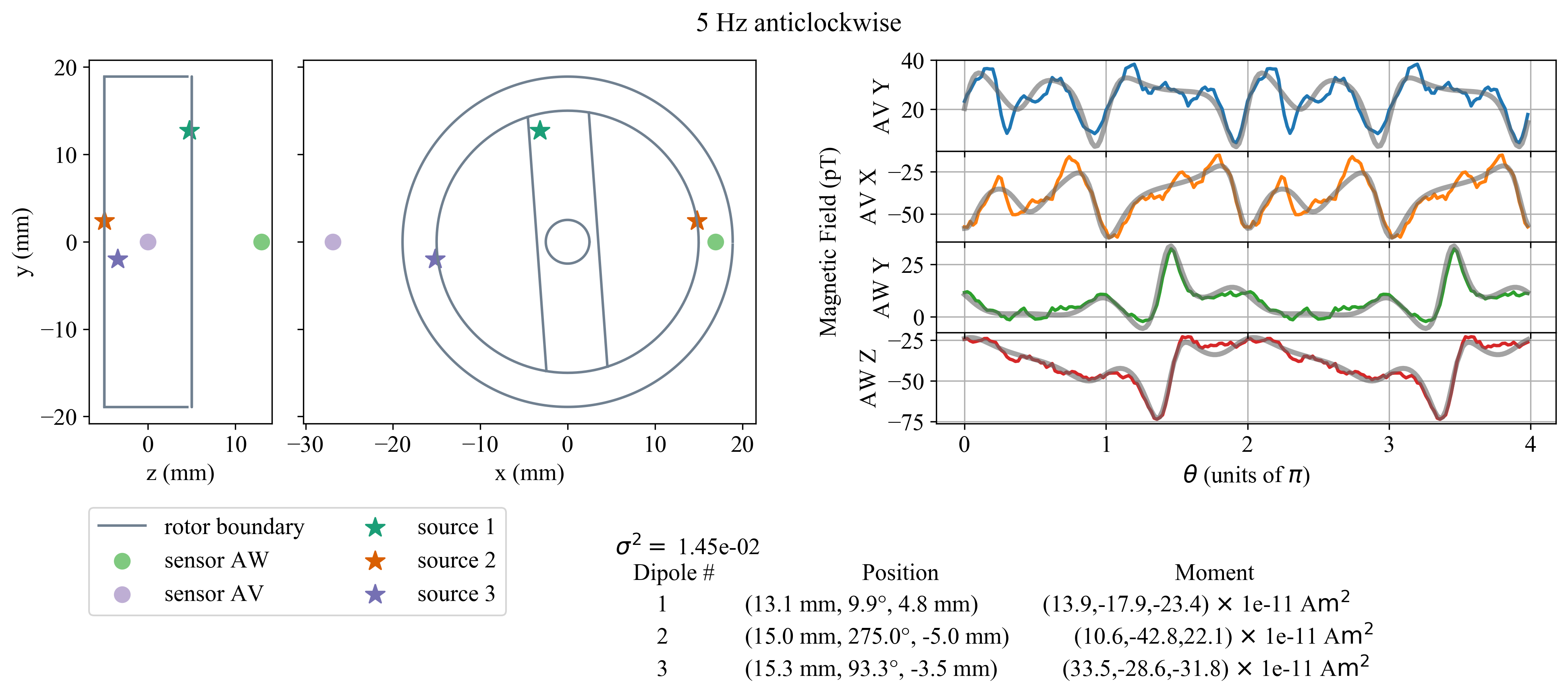}
 	\caption{Fitting Results: On the right, we show the recorded magnetic field data (see text for description of data processing) from each sensor in colored lines, and the magnetic field produced by three hypothetical dipoles in gray lines. This is shown as a function of \(\theta_\mathrm{rot}\), the angle by which the rotor is rotated.  On the left, we show a schematic of the problem: a front and side view of the rotor with each sensor's location indicated with a circle. The dipole locations and moments found by the fitting algorithm are listed at the bottom. The locations are also indicated in the schematic as stars.}
 	\label{fig:fitfig}
 \end{figure*}

 For all other frequencies, since the time-domain measurements line up, we can focus on analyzing any one of them. In this paper we will focus on \SI{5}{Hz} since it's the most relevant for the axion measurement\opendata{
\footnote{the analysis for all other frequencies can be found in the data and code released with this publication}}. 
 We average the measurement over a few cycles, the averaged data is shown by the colored curves on the right side of \cref{fig:fitfig}. 
 
 We now try to answer whether this measurement can be explained by a few discrete dipoles. For three dipoles, the problem is parametrized into 23 free parameters: the position and magnetic moments of each dipole \((6\times N_\mathrm{dipoles})\), DC offsets in the sensors \((N_\mathrm{sensors}\times N_\mathrm{axes})\), and the initial location of the horizontal bar with respect to the x-axis in the lab frame \((\theta_\mathrm{bar})\).

 Given the field from a magnetic dipole
\begin{equation}
	\mathbf{B} = \frac{\mu_0}{4\pi} \left(3\frac{\mathbf{m}\cdot\mathbf{r}}{r^5}\mathbf{r}-\frac{\mathbf{m}}{r^3}\right),
\end{equation}
for a source inside the rotor, rotation varies the position of the source \(\mathbf{r}(t)\), giving a time-varying field \(\mathbf{B}(t)\).

For each sensor and source, we define the problem in cylindrical coordinates as follows:

\begin{subequations}
\begin{align}
\mathbf{r}(t) &= \mathbf{R}_\mathrm{OPM} - \mathbf{R}_\mathbf{src}(t)\\
	\mathbf{m} &= (m_r,m_\theta,m_z),\\
	\mathbf{R}_\mathrm{src} &= (R_\mathrm{src},\theta_\mathrm{total}(t),z_\mathrm{src})\\
	\theta_\mathrm{total}(t) &= \theta_\mathrm{bar} + \theta_\mathrm{src} + \theta_\mathrm{rot}(t),\\
	\theta_\mathrm{rot}(t) &= \omega_\mathrm{rot}(t-t_0) \label{eq:thetarot}
\end{align}
\end{subequations}

For a particular source moment and location as well as a known rotation speed, a vector magnetic field at the location of each OPM can be calculated using the above prescription. The total field at the location of $j$-th OPM will be the sum of fields from all dipoles:

\begin{equation}
\mathbf{B}_j^\mathrm{model}(t) = \sum_{i=1}^{N_\mathrm{dipoles}} \mathbf{B}_{ij}(t)
\end{equation}
Then a normalized error function is defined:
\begin{equation}
\sigma^2 = \frac{1}{N^2}\sum_{j,k,t}\left(\frac{
{B}_{jk}^\mathrm{model}(t)-{B}_{jk}^\mathrm{meas}(t)}{\Delta B}
\right)^2
\end{equation}
where \(N\) is the total number of data points, \(\Delta B\) is the measurement uncertainty (taken to be \SI{1}{\pico\tesla}), $j$ runs over the two OPMs (AV and AW), and $k$ runs over the two measurement axes of each OPM.

We then use the Nelder-Mead numerical algorithm \cite{gao2012implementing} implemented by the Python function scipy.optimize.minimize() to find the parameters that minimize the above error function. The result of the optimization is shown in \cref{fig:fitfig}. We show the fitted locations of the impurities in a schematic of the rotor, with their magnetic moments printed below. The field from these dipoles is shown in comparison with the measured field.
 The moments of the dipoles that minimize the $\sigma^2$ have been bounded to be less than \Number{\SI{1e-9}{\ampere\meter\squared}}.


\section{Conclusion}

This work characterizes the noise and background caused by the source mass in the ARIADNE experiment. This source would be at the same frequency as the signal and would be completely indistinguishable from the signal, hence the only way forward is to reduce it below the expected signal level.

We measured the magnetic field in proximity to the ARIADNE rotor after demagnetizing it. We find that the contribution of magnetic field amplitude at the Larmor frequency (which is 11 times the rotation frequency) is below \SI{1}{\pico\tesla} and meets the design requirement. To understand the mechanism producing the field, we also model the measurement as magnetic field created by three magnetic dipoles. We find good agreement between the measurement and the model, supporting the hypothesis that a significant portion of the measured field is created by trapped magnetic impurities inside the rotor. In the future, if this becomes a limiting background, a purer material or better demagnetization techniques can be sought.

   Going forward, the rotation speed can be tracked by following the peaks in any over time in any one of the four channels. A separate channel that measures speed can also be deployed. An interferometric readout of the rotation speed is planned for the ARIADNE experiment \cite{Ariadneproposal}. This will allow for averaging over longer time as well as a more precise estimate of the contribution from Johnson noise vs coherent magnetic field background. 

While the Nelder-Mead method is adequate to provide order of magnitude estimates of the dipole moments of trapped impurities and validates the hypothesis that a significant portion of the magnetic field noise is caused by a few discrete impurities, this method does have a number of limitations. For example, there are many degeneracies in the parameter space, making the method sensitive to the initial guess given to the optimizer. Multi-modal solutions can be missed since the approach only provides one local minimum
     	. In future, a Markov-Chain Monte-Carlo method based approach may provide results independent of the initial guess and better estimate the uncertainty in the magnitude and locations of magnetic impurities. Data with additional magnetometers at more spatial locations around the rotor could also help to improve the estimate of location and moments of the impurities.

\section{Acknowledgments}

We acknowledge support from the U.S. National Science Foundation, grant numbers NSF PHY-1509176, NSF PHY-1510484, NSF PHY-1506508, NSF PHY-1806671, NSF PHY-1806395, NSF PHY-1806757.

A. Brown, I. Lee, C.-Y. Liu, J. C. Long, A. Reid, J. Shortino, E. Smith, and W. M. Snow acknowledge support from the Indiana University Center for Spacetime Symmetries. 

We also acknowledge funding support from IBS, grant number IBS-R017-D1-2020-a00.

We acknowledge the 
support of the Core Facility 'Metrology of Ultra-Low Magnetic Fields' at Physikalisch-Technische 
Bundesanstalt which receives funding from the Deutsche Forschungsgemeinschaft – DFG (funding 
codes: DFG KO 5321/3-1 and TR 408/11-1).

\section{Author Contributions}

AB and JCL fabricated the source mass. AS and JV performed the measurements, NA analyzed the data and led this manuscript.  All authors contributed to the final manuscript. 

\opendata{\section{Data and Code Availability}

The raw data as well as the code to perform all the analysis needed for this manuscript can be downloaded from \cNA{to-be-decided}
\todo{Create a zenodo/Github for this}}

\bibliography{references}
\clearpage

\end{document}